\documentclass[trackchanges]{aastex701}

\begin{document}

\title{Ocean Circulation on Tide-locked Lava Worlds: 3D Modeling with a Simple Boundary Iteration Method}

\renewcommand{\thefootnote}{\fnsymbol{footnote}}


\author[0000-0001-6031-2485]{Jun Yang}
\affiliation{Laboratory for Climate and Atmosphere-Ocean Studies, Dept. of Atmospheric \& Oceanic Sciences, School of Physics, Peking University, Beijing 100871, China.}
\affiliation{Institute of Ocean Research, Peking University, Beijing 100871, China}
\email{junyang@pku.edu.cn}

\author{Chengyao Tang}
\affiliation{Laboratory for Climate and Atmosphere-Ocean Studies, Dept. of Atmospheric \& Oceanic Sciences, School of Physics, Peking University, Beijing 100871, China.}
\email{2200011416@stu.pku.edu.cn}


\author{Zimu Wang}
\affiliation{Laboratory for Climate and Atmosphere-Ocean Studies, Dept. of Atmospheric \& Oceanic Sciences, School of Physics, Peking University, Beijing 100871, China.}
\email{m15156475611@stu.pku.edu.cn}

\author[0000-0001-9700-9121]{Yanhong Lai}
\affiliation{Tsung-Dao Lee Institute, Shanghai Jiao Tong University, 1 Lisuo Road, Shanghai 200127, China}
\email{yanhonglai@sjtu.edu.cn}

\author[0000-0002-4615-3702]{Wanying Kang}
\affiliation{Earth, Planetary \& Atmospheric Sciences, MIT, MA 02139, USA}
\email{wanying@mit.edu}

\footnotetext[1]{Corresponding Author: Jun Yang, Email: junyang@pku.edu.cn}

\begin{abstract}
Tide-locked lava worlds are surface-melted rocky planets under 1:1 tidally locked orbit \textcolor{black}{(i.e., synchronously rotating)} with orbital period being equal to rotation period and with permanent hot dayside and cold nightside. Previous studies on this type of planets employed scaling analyses and two-dimensional (2D) simulations. This work is a continuation of the previous researches but including the effect of the Coriolis force and the simulation domain is extended to a 3D global sphere. We find that under the condition with thermal-only forcing (without surface wind stresses)\textcolor{black}{, the area-mean ocean depth is about 50--300 m (depending on vertical diffusivity) and the area-mean effect of horizontal ocean heat transport \textcolor{black}{(in the order of 10$^{3}$ to 10$^{4}$\, W\,m$^{-2}$)} is significantly smaller than stellar radiation \textcolor{black}{(in the order of 10$^{6}$\, W\,m$^{-2}$ at the substellar region)}}, being consistent with previous results. Different from 2D results, due to the effect of the Coriolis force, large-scale horizontal gyres form on the dayside, ocean currents near the west boundaries are much stronger than that near the east boundaries (called as ``western intensification''), the deepest ocean is not right at the substellar point but in the middle latitudes \textcolor{black}{as the vertical diffusivity is moderate or large}, and meanwhile there exists significant asymmetry between the west and the east of the substellar point. These results establish a first picture for the 3D thermal-driven ocean circulation and confirm that the lava ocean should be shallow on tide-locked lava worlds.
\end{abstract}



\section{Introduction} \label{sec:intro}

Among the more than 5800 exoplanets discovered to date, tens to hundreds are likely to be 1:1 tidally locked \textcolor{black}{(i.e., synchronously rotating)} lava worlds, such as TOI-561b, K2-141b, Kepler-10b, 55 Cnc e, and CoRoT-7b, particularly if their orbital eccentricities are close to zero \citep[e.g.,][]{leger2009, leger2011, chao2021}. \textcolor{black}{If the orbital eccentricity is moderate or large, the planet will likely enter into other types of tidally locked state such as spin-orbit resonance, similar to the planet of Mercury, of which the ratio of spin period to orbital period is 2:3. In this study, we focus on the synchronously rotating state.} In the context of ocean fluid dynamics, several key questions arise for these tidally locked lava planets: How wide is the ocean in the horizontal direction? How deep is the ocean? How efficient is the horizontal ocean heat transport? Previous studies on this topic have reached two opposite conclusions.

On one hand, if strong internal heating is present at the ocean bottom due to delayed differentiation or radioactive decay in early formation processes, the entire ocean may become convective. In this case, the ocean depth is determined by the intersection between the liquidus curve (i.e., the melting point of rocks as a function of pressure) and the adiabatic temperature profile (set by vigorous convection). As a result, the ocean depth can reach 100 km or more \citep[e.g.,][]{leger2011, nguyen2020, Boukare2022, Meier2023}.
An analogy can be drawn with Earth's atmosphere, where convection penetrates the entire troposphere, inducing efficient mixing, because most of solar radiation reaches \textcolor{black}{the surface} (the bottom of the atmosphere).

On the other hand, if internal heating is weak, the ocean depth is instead controlled by the efficiency of ocean circulation, which transports heat downward through mean flow and mixing processes. In this condition, the ocean remains relatively shallow. Theoretical scaling analyses and two-dimensional (2D) numerical simulations suggest ocean depths on the order of 100–1000 m \citep{Kite2016, lai2024a, lai2024b}.
Considering that tidally locked lava planets have cooled several billion years and that their oceans are predominantly heated from above by stellar irradiation rather than from below by internal heat, a shallow ocean appears to be the more plausible scenario.

In the absence of the Coriolis force, idealized 2D models estimate horizontal current speeds based on a balance between pressure gradient and viscous terms \citep{lai2024a}, which cannot realistically simulate the strength of ocean currents and the associated horizontal heat transport. While previous 2D models captured only zonal and vertical currents, meridional heat transport is expected to play an important role, as is well established for  Earth's oceans \citep{vallis2017}. The inclusion of planetary rotation alters the direction of zonal flows through the Coriolis effect—deflecting to the right in the northern hemisphere and to the left in the southern hemisphere. To address these limitations, we extend previous 2D modeling to full 3D simulations that incorporate the Coriolis effect. The model setup is described in Section~\ref{sec:method}, followed by the main results and comparisons with scaling theory in Section~\ref{sec:results}. Summary and discussion are presented in Section~\ref{sec:summary}.


\section{Method} \label{sec:method}

\subsection{Global Ocean Modeling}
We employ the MIT General Circulation Model (MITgcm; \citealt{marshall1997_a, marshall1997_b}) with moderate modifications to simulate the magma ocean dynamics of tidally locked lava planets. The model solves the primitive equations in spherical coordinates with a free surface. The Coriolis force is included and expressed as $2\Omega \sin\theta$, where $\Omega$ is the planetary rotation rate and $\theta$ is latitude. For simplicity, only the liquid (magma ocean) component is simulated, while the underlying and lateral solid region is assumed to be motionless. The experimental setup is primarily based on MITgcm’s online Experiment 4.5\footnote{\url{https://mitgcm.readthedocs.io/en/latest/examples/global_oce_latlon/global_oce_latlon.html}}.

The ocean circulation is not dynamically coupled to an atmosphere, but is instead driven by an imposed thermal forcing. Specifically, the sea surface temperature is relaxed toward a prescribed radiative equilibrium distribution with a relaxation timescale of 30 Earth days. \textcolor{black}{It is estimated based on the equation of $\tau=C_p\rho H/(4\sigma T_e^3)$, where $H$ is the mixed layer depth of the ocean, $C_p$ is specific heat capacity, $\rho$ is lava density, $\sigma$ is the Stefan-Boltzmann constant, and $T_e$ is effective emission temperature. Given $C_p=3994$\,J\,kg$^{-1}$K$^{-1}$, $\rho=2673$\,kg\,m$^{-3}$, $H=100$\,m, $\sigma=5.67\times10^8$\,W\,m$^{-2}$\,K$^{-4}$, $T_e=1000$\,K, the value of $\tau$ is about 55 Earth days. For an emission temperature is 3000~K, the value of $\tau$ decreases to 1.8 Earth days. We use a value between these two numbers, a quick method but not perfect. The atmosphere is likely thin (less than 0.1~bar) because of significant long-term atmospheric escape \citep[e.g.,][]{SchaeferandFegley2009}, so its thermal inertia is not considered in this estimate.} 

The target temperature pattern follows a cosine function in both latitude and longitude, peaking at 3000 K at the substellar point and decreasing to a uniform 50 K on the nightside (equivalent to a uniform geothermal heat flux of 0.35 W\,m$^{-2}$). Surface wind stresses are not included in this work, but will be considered in next step of this series of studies.

We adopt planetary parameters representative of Kepler-10b \citep{batalha2011, dumusque2014}. Planetary radius is $9.4\times10^{6}$ m, orbital period is 0.83 Earth days (being equal to the rotation period under synchronous rotation), and surface gravity is 14.7 m\,s$^{-2}$. The incident stellar radiation at the substellar point is $4.59\times10^6$ W m$^{-2}$, corresponding to a substellar surface temperature of 3000 K assuming zero planetary albedo. The physical properties of molten lava ocean include thermal expansion coefficient of $2 \times 10^{-4}$ K$^{-1}$, specific heat capacity of 3994 J\,kg$^{-1}$\,K$^{-1}$, density of 2673 kg\,m$^{-3}$, and freezing point of 2000 K. \textcolor{black}{These values are set following laboratory measurements of \citet{Katsuraetal2010} and \citet{Sakamakietal2010}. The exact values should depend on pressure, temperature, lava compositions and other factors, but it is unlikely to influence the main conclusion of this study.} 


Turbulent viscosity and diffusivity play critical roles in shaping ocean circulation. In the liquid domain, we apply constant vertical and horizontal viscosity coefficients of $4 \times 10^{-2}$ and $2 \times 10^4$ m$^2$\,s$^{-1}$, respectively. Vertical and horizontal thermal diffusivities are similarly fixed at $4 \times 10^{-5}$ and $4 \times 10^3$ m$^2$\,s$^{-1}$. These values are somewhat higher than typical values for Earth's seawater, yet remain within the same order of magnitude. Previous experimental studies have shown that fully molten silicates can exhibit viscosity and diffusivity comparable to those of seawater \citep{dingwell2004, sun2020, zhang2022}. \textcolor{black}{Their values are roughly set to be close to the mean value of Earth’s seawater ocean whereas their exact values can only be obtained through very high-resolution simulations, such as 10 or 100 m, which is beyond the ability of present computation power. Note that these values are for eddy diffusivity and viscosity, rather than molecular diffusivity and viscosity. The horizontal viscosity or diffusivity is much larger than that for vertical viscosity or diffusivity because the horizontal scale (100-1000 km) is much greater than the vertical scale (100-1000 m) and the coefficients should be proportional to the spatial scale. Moreover, the vertical velocity of turbulence and eddies is always weaker than the corresponding horizontal velocity. For more details, please see the discussions in section 2.1 of \citet{lai2024a}}. Due to the extreme contrast between the liquid and solid silicate rheology, we omit the solid component in our simulations. Future high-resolution, eddy-resolving models may provide better constraints on these unresolved parameters.

The model domain spans from $80^\circ$S to $80^\circ$N to avoid computational singularities near the poles (which are not supported in MITgcm). The horizontal resolution is $2^\circ$ in both latitude and longitude, corresponding to 80 meridional grid cells and 180 zonal grid cells. The vertical resolution is variable, with finer resolution (16 m) near the surface and coarser resolution at depth. All simulations are integrated using a constant time step of 300 s.

\subsection{Boundary Iteration Method and Its Sensitivity}
\begin{figure*}[ht!]
    \centering
    \includegraphics[width=\textwidth]{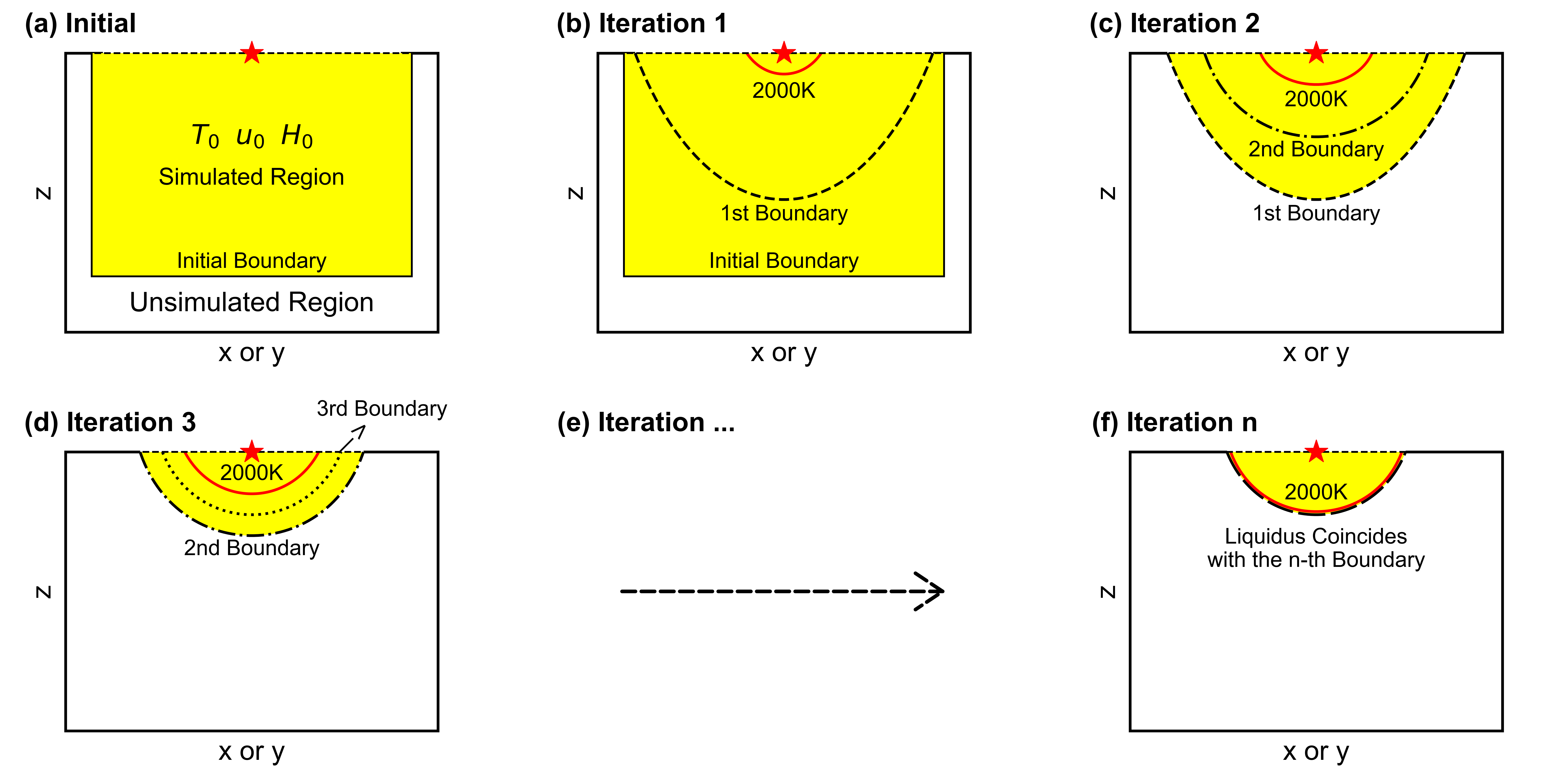} 
    \caption{
        Schematic diagram of the boundary iteration method. The red star marks the substellar point. In each panel, the colored region represents the active simulation domain, while the white area is not simulated. Solid and dashed black lines denote the simulation boundaries, and the red curve indicates the 2000 K isotherm (freezing point). At each iteration, the new simulation boundary is defined as the midpoint between the 2000 K isotherm and the previous iteration's boundary. Iterations continue until the simulation boundaries converge. Initial lateral and bottom ocean boundaries are arbitrarily defined (see Figure~\ref{fig2}).
    }
    \label{fig1}
\end{figure*}

A key challenge in simulating tidally locked lava oceans lies in the fact that their lateral and bottom boundaries are unkown. In our previous 2D simulations of \cite{lai2024a}, we explicitly modeled all three phases—liquid, partially molten, and solid—by allowing the viscosity to vary strongly with temperature. This approach, however, led to extreme viscosity contrasts (spanning orders of magnitude) that introduced significant numerical instabilities. To address this, \citet{lai2024a} developed a backward implicit time-stepping scheme to stabilize the primitive equations. In contrast, the standard MITgcm setup employed in this study uses an explicit time integration scheme and cannot stably simulate such large viscosity contrasts. Therefore, in the present work, we limit our focus to the liquid phase only. This introduces a new problem: the boundaries of the liquid magma ocean are not fixed but evolve dynamically in response to the circulation itself, which in turn is influenced by these boundaries. To address this coupling, we develop a simple but effective boundary iteration method, schematically illustrated in Figure~\ref{fig1}.

In this method, the model is first initialized with a large ocean domain filled with liquid lava (Figure~\ref{fig1}(a)). The fluid is assigned with standard viscosity and diffusivity values, and optionally, strong initial currents to maximize the initial heat transport. Once the simulation reaches equilibrium in both temperature and velocity fields, we identify the 2000 K isotherm and use it to redefine the simulation domain (Figure~\ref{fig1}(b)). Specifically, we scan the model grid horizontally and vertically to locate the 2000 K isotherm, and then construct new lateral and bottom boundaries at the midpoint between this isotherm and the previous simulation boundary (black dashed lines in Figure \ref{fig1}(b)). This newly defined boundary is then used to reinitialize the simulation, and the process is repeated; \textcolor{black}{the model is run until equilibrium or quasi-equilibrium is reached,} the 2000 K isotherm is extracted, and the simulation domain is further updated (Figure~\ref{fig1}(c-e)). This iterative process continues until convergence is achieved---that is, \textcolor{black}{any point along the 2000 K contour line contacts the simulation boundary, rather than requiring the entire 2000-K contour to coincide with the boundary, due to computation time limitations} (Figure~\ref{fig1}(f)). Typically, rough convergence is reached after 7 to 10 iterations, with each iteration spanning approximately 300 model years or more.

\begin{figure*}[ht!]
    \centering
    \includegraphics[width=\textwidth]{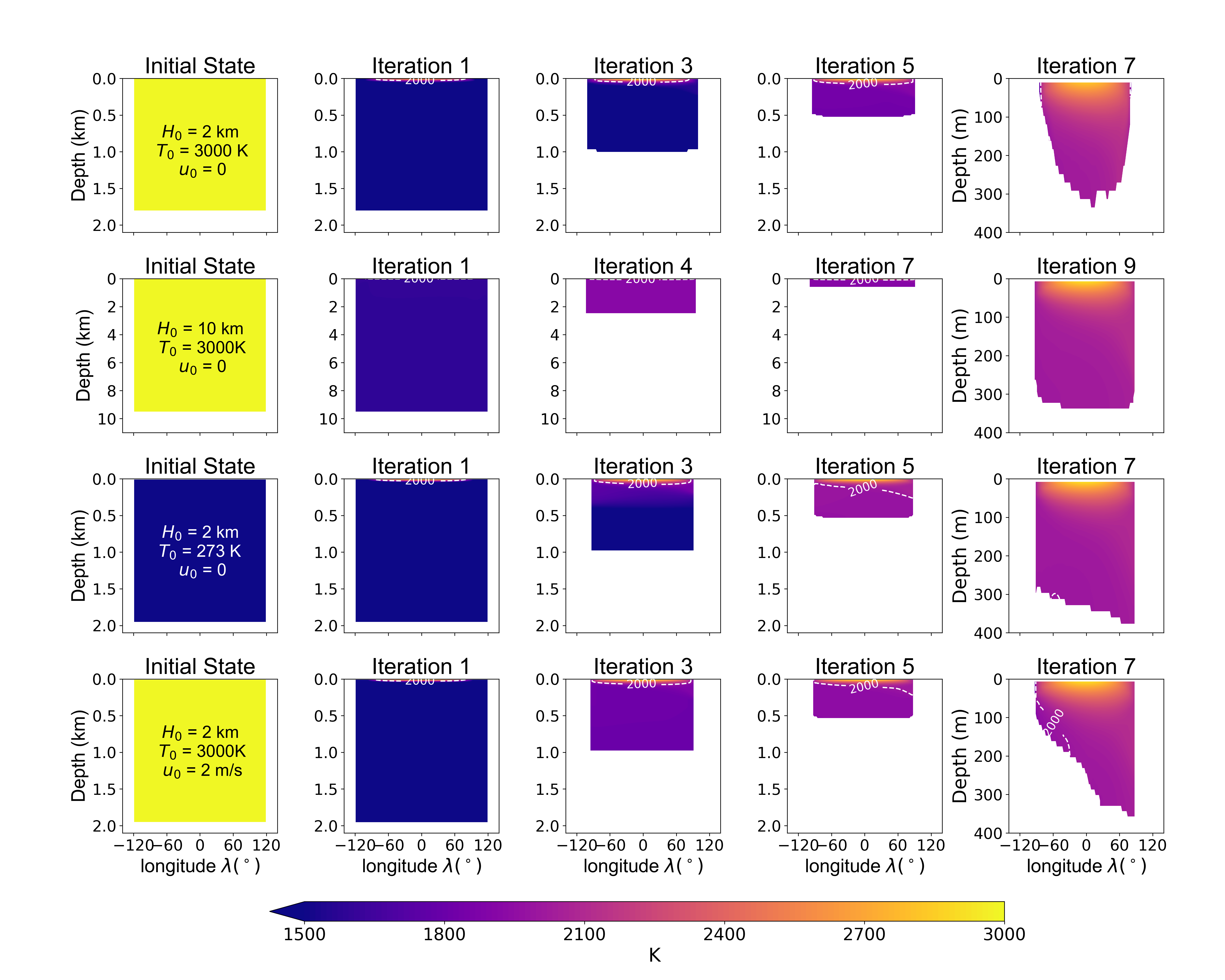} 
    \caption{
        Sensitivity tests of the boundary iteration method under four different initial conditions. The top row shows the control case, and the other rows show variations in initial ocean temperature, depth, and current strength. The first column displays the initial condition for each test, while the remaining columns show intermediate and final states during the iteration process. The colored regions represent the active simulation domain; white areas are not simulated. Although the initial conditions vary significantly, all experiments \textcolor{black}{roughly} converge to similar equilibrium states with an ocean depth of approximately 200–400 m along the equator. Note: The y-axis range differs in the rightmost column.
    }
    \label{fig2}
\end{figure*}

To assess the robustness of this method to different initial conditions, we performed four sensitivity experiments (Figure~\ref{fig2}) by varying three key parameters: the initial depth of the ocean ($H_0$), the initial temperature field ($T_0$), and the strength of the initial currents ($u_0$). The first column shows the initial conditions for each case, while subsequent columns display equatorial cross sections of lava temperature across successive iterations. In each panel, the white dashed lines indicate the 2000 K isotherm. \textcolor{black}{The last column shows the final quasi-equilibrium state. There are significant differences in the quasi-equilibrium state under the four different initial states: the ocean in the west part of the substellar point is shallower than in the east part \textcolor{black}{in all the four experiments}; the west-to-east asymmetry in the experiments with an initial depth of 2 km is larger than that in the experiment of 10 km; and the largest asymmetry occurs in the experiment with an initial ocean depth of 2 km and meanwhile with an initial uniform ocean current of 2 m\,s$^{-1}$. This  west-to-east asymmetry is due to that there is strong downwelling motion near the east boundary of the ocean domain (see Figure~\ref{fig4}c below), which transports hot lava from the surface to the ocean bottom. In the experiment with an initial ocean current of 2 m\,s$^{-1}$ (due to an artifical error, this initial current was added to \textcolor{black}{the beginning of} each iteration of this experiment), this strong current enhances this asymmetry, so that its asymmetry is the largest among the four experiments. Despite the dependence of the final equilibrium state on the initial state, all the four experiments have similar area-mean ocean depth, around 250 to 400 m along the equator (the rightmost column in Figure~\ref{fig2}).} This suggests that while the method is simple and not fully self-consistent in simulating the dynamics of solidification, it  \textcolor{black}{can roughly} capture the primary feature of magma ocean circulation and determine the spatial domain of the liquid phase.


\section{Results} \label{sec:results}

\begin{figure*}[ht!]
    \centering
    \includegraphics[width=\textwidth]{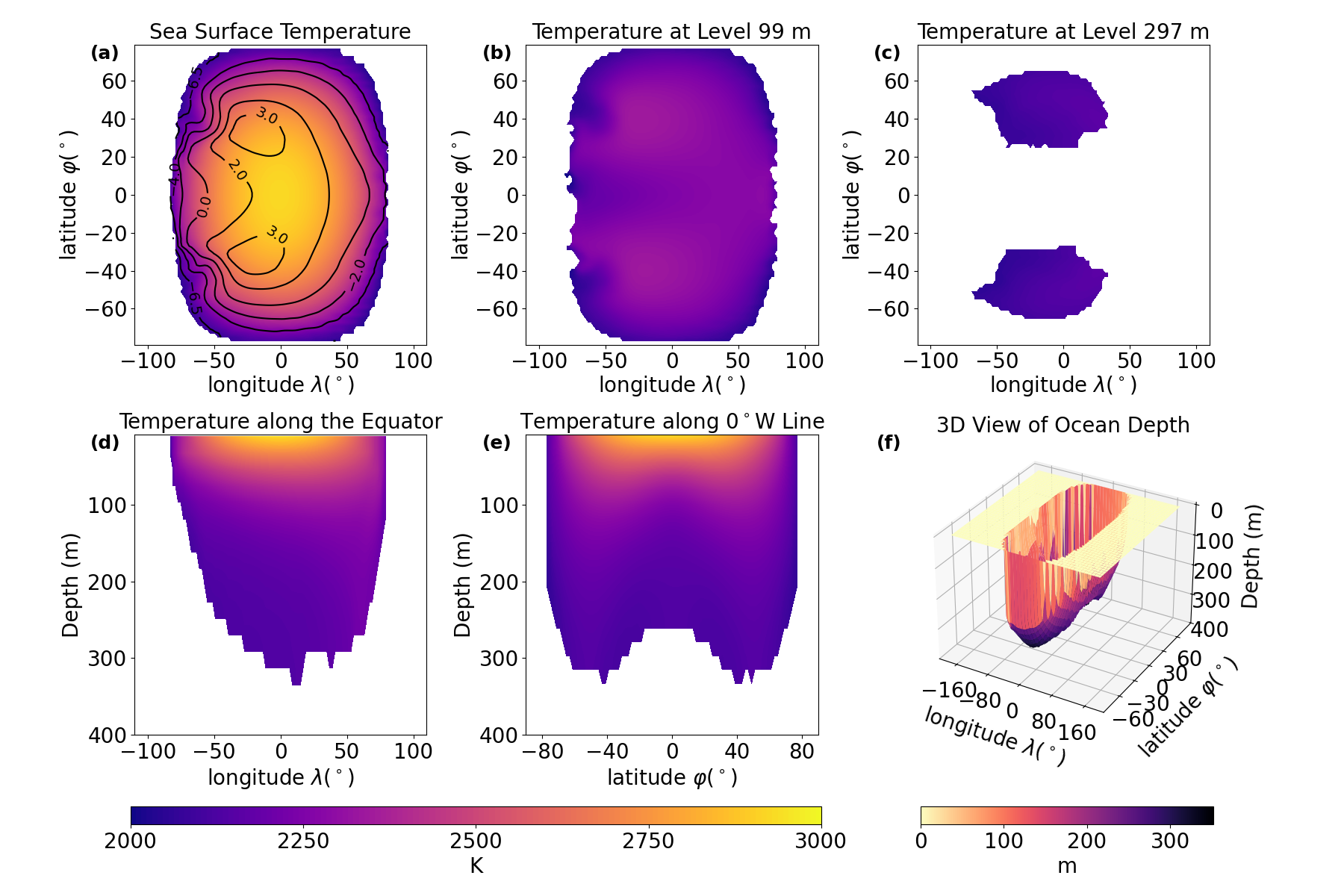} 
    \caption{
        Lava temperature and ocean depth in the control experiment with a vertical diffusivity of $4 \times 10^{-5}$ m$^2$ s$^{-1}$. (a) Sea surface temperature \textcolor{black}{(K, color-shaded)} and surface surface height (\textcolor{black}{m, contour lines}); (b, c) temperature slices at 99 m and 297 m depth; (d) equatorial cross-section temperature; (e) meridional cross-section temperature at the substellar longitude ($0^\circ$); (f) 3D perspective view of ocean depth (m). \textcolor{black}{On the night side and near the terminators, the ocean depth is zero everywhere.} White areas indicate solid regions (temperature below 2000 K), which are not included in the simulation domain (the same applies to all following figures). 
    }
    \label{fig3}
\end{figure*}

Figure~\ref{fig3} presents the distribution of ocean temperature and ocean depth in the control simulation with a vertical diffusivity of $4\times10^{-5}$ m$^{2}$\,s$^{-1}$. The maximum sea surface temperature reaches approximately 2926 K, located near the substellar point, and gradually decreases with distance from that point. Near the edges of the ocean, the lava temperature approaches 2000 K (i.e., the prescribed melting point), beyond which the lava solidifies and is excluded from the simulation domain. Horizontally, the ocean extends approximately $84^\circ$ westward and $82^\circ$ eastward from the substellar point.
The ocean depth ranges from roughly 100 to 400 m, with a maximum depth of about 350 m. This is comparable to the depth of the thermocline in Earth's oceans, which marks the level with the sharpest vertical gradient in seawater temperature and density, beneath which stratification is much weaker \citep{vallis2017}.

While the northern and southern hemispheres are nearly symmetric, there is a noticeable asymmetry between the western and eastern hemispheres. This asymmetry arises due to the influence of planetary rotation. On a rotating sphere, the intense horizontal temperature gradient induced by stellar heating excites tropical wave modes: Rossby waves that propagate westward and Kelvin waves that propagate eastward. This pattern is similar to the wave-driven responses in Earth’s equatorial oceans and in the atmospheres of tidally locked exoplanets \citep[e.g.,][]{stewart2008, showman2008, showman2013, zeng2021}.
As shown in Figure~\ref{fig3}(e), the deepest part of the magma ocean is located at approximately $40^\circ$S/N, rather than at the equator. This latitudinal shift is attributed to poleward heat transport associated with the Rossby wave dynamics. In both hemispheres, the ocean bottom forms a bowl-like structure (Figure~\ref{fig3}(f)).

\begin{figure*}[ht!]
    \centering
    \includegraphics[width=\textwidth]{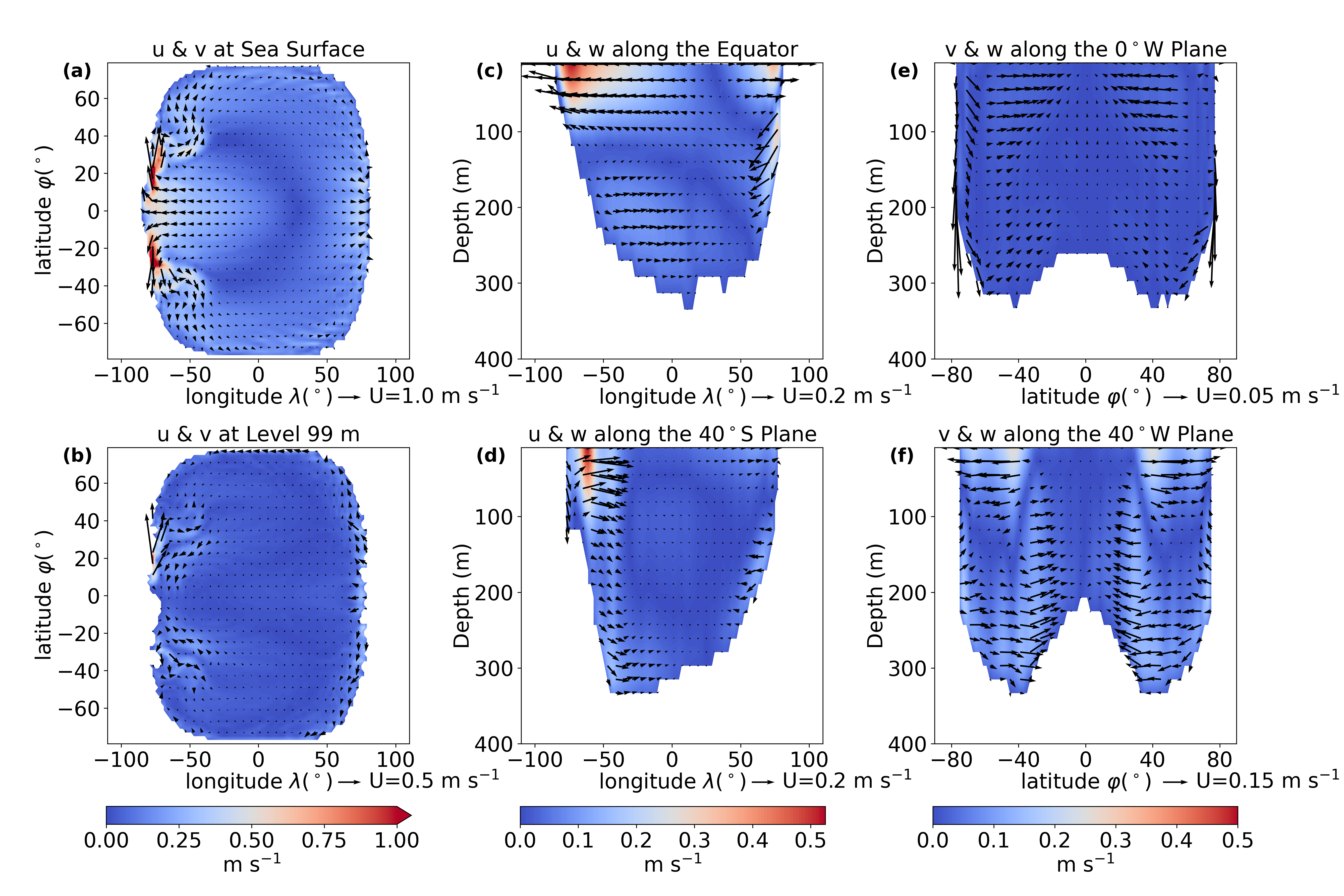} 
    \caption{
        Ocean currents in the control experiment. (a) Horizontal currents ($u$, $v$) at the sea surface; (b) horizontal currents at 99 m depth; (c) zonal-vertical section ($u$, $w$) along the equator; (d) zonal-vertical section at $40^\circ$S; (e) meridional-vertical section ($v$, $w$) along $0^\circ$ longitude; (f) meridional-vertical section along $40^\circ$ longitude, in the west of the substellar point. Note that the color bar range (representing velocity magnitude, $\sqrt{u^2 + v^2 + w^2}$) and the reference vector length vary between different panels.
    }
    \label{fig4}
\end{figure*}

Figure~\ref{fig4} shows the 3D structure of ocean currents across various layers and cross sections. At the surface, horizontal flow velocities are on the order of 0.1–1.0 m\,s$^{-1}$ and decrease with depth. Vertical velocities are much smaller, ranging from $10^{-6}$ to $10^{-5}$ m\,s$^{-1}$. This difference reflects the strong aspect ratio of the system---horizontal length scales on the order of $10^7$ m compared to vertical scales of $\sim10^2$ m, giving an aspect ratio near $10^5$.

Scaling analyses indicate that the dominant balance in the horizontal momentum equation is between the Coriolis force and the pressure gradient, both of which are on the order of $\sim\,10^{-5}$ m\,s$^{-2}$. This suggests that the horizontal currents are primarily in geostrophic or quasi-geostrophic balance. Other terms are at least two orders of magnitude smaller; for instance, the vertical viscosity term is on the order of $\sim\,10^{-7}$ m\,s$^{-2}$, while both vertical and horizontal advection terms are approximately $\sim\,10^{-9}$ m\,s$^{-2}$. The Rossby number, defined as the ratio of inertial (advection) to Coriolis forces, is on the order of $10^{-4}$ for large-scale circulations, further confirming the geostrophic nature of the flow. In the temperature equation, the dominant terms include vertical and horizontal advections, as well as vertical diffusion, each on the order of $\sim\,10^{-7}$ K\,s$^{-1}$. By contrast, horizontal diffusion is about three orders of magnitude smaller and thus negligible in comparison.

Due to thermal expansion, \textcolor{black}{there is a local maximum point in sea surface height (SSH) around the substellar point} (see contour lines in Figure~\ref{fig3}(a)), where the sea surface temperature peaks (color-shaded). This SSH distribution sets up a westward pressure gradient on the west side of the substellar point, driving westward surface currents, and an eastward pressure gradient on the east side, driving eastward surface currents (Figure~\ref{fig4}(a)). The Coriolis force deflects the westward currents to the right in the northern hemisphere and to the left in the southern hemisphere. As a result, surface currents form a large-scale anticyclonic (clockwise) gyre in the northern hemisphere and a cyclonic (counterclockwise) gyre in the southern hemisphere, each spanning the scale of the entire ocean basin (~$10^7$ m). \textcolor{black}{Note that the maximum value in SSH is not right on the equator but in the latitude of 40$^o$S(N), being consistent with the deepest ocean there.}

In this 3D simulation, the magnitude of SSH is several meters (Figure~\ref{fig3}(a)), being about one order lower than that in previous 2D simulations (see Figure 4(d) in \cite{lai2024a}). This is likely due to the fact that the meridional heat transport in the 3D frame (but absent in the 2D frame) acts to weaken the thermal contrast in the zonal direction along the equator. \textcolor{black}{The 3D frame effectively transports heat from the equatorial region to the extra-tropical regions.}

As shown in Figure~\ref{fig4}(a)–(b), surface currents are significantly stronger and more concentrated along the western boundaries of the ocean basin than along the eastern boundaries. This pattern resembles the phenomenon of western intensification, widely observed in Earth's Pacific and Atlantic Oceans. This intensification arises from the beta effect \citep{stommel1948, munk1950, vallis2017}, i.e., the latitudinal variation of the Coriolis parameter, which is inherently captured in the present 3D simulations but absent in previous 2D models. In the ocean interior, the broad, weak equatorward flow is compensated by a narrow, intense poleward return flow along the western boundary.

In the vertical direction, the vertical velocity is small but still exhibit clear spatial patterns (Figure~\ref{fig4}(c)–(f)). The lava ocean features widespread upwelling across much of the basin, particularly near the substellar region, while the regions near the lateral and bottom boundaries of the ocean are dominated by downwelling. In the substellar region, fluid is heated and rises toward the surface, increasing the system’s gravitational potential energy. Meanwhile, fluid near the ocean boundaries cools and sinks, releasing potential energy. The vertical motions near the ocean edges are notably stronger than those in the interior, consistent with the pattern of Earth’s ocean circulation \citep[e.g.,][]{boyd2014, vallis2017}. Together with horizontal currents, these vertical motions establish a global 3D overturning circulation: upwelling at the substellar point, lateral transport toward the boundaries, downwelling near the edges, and return flow at depth.

\begin{figure*}[ht!]
    \centering
    \includegraphics[width=\textwidth]{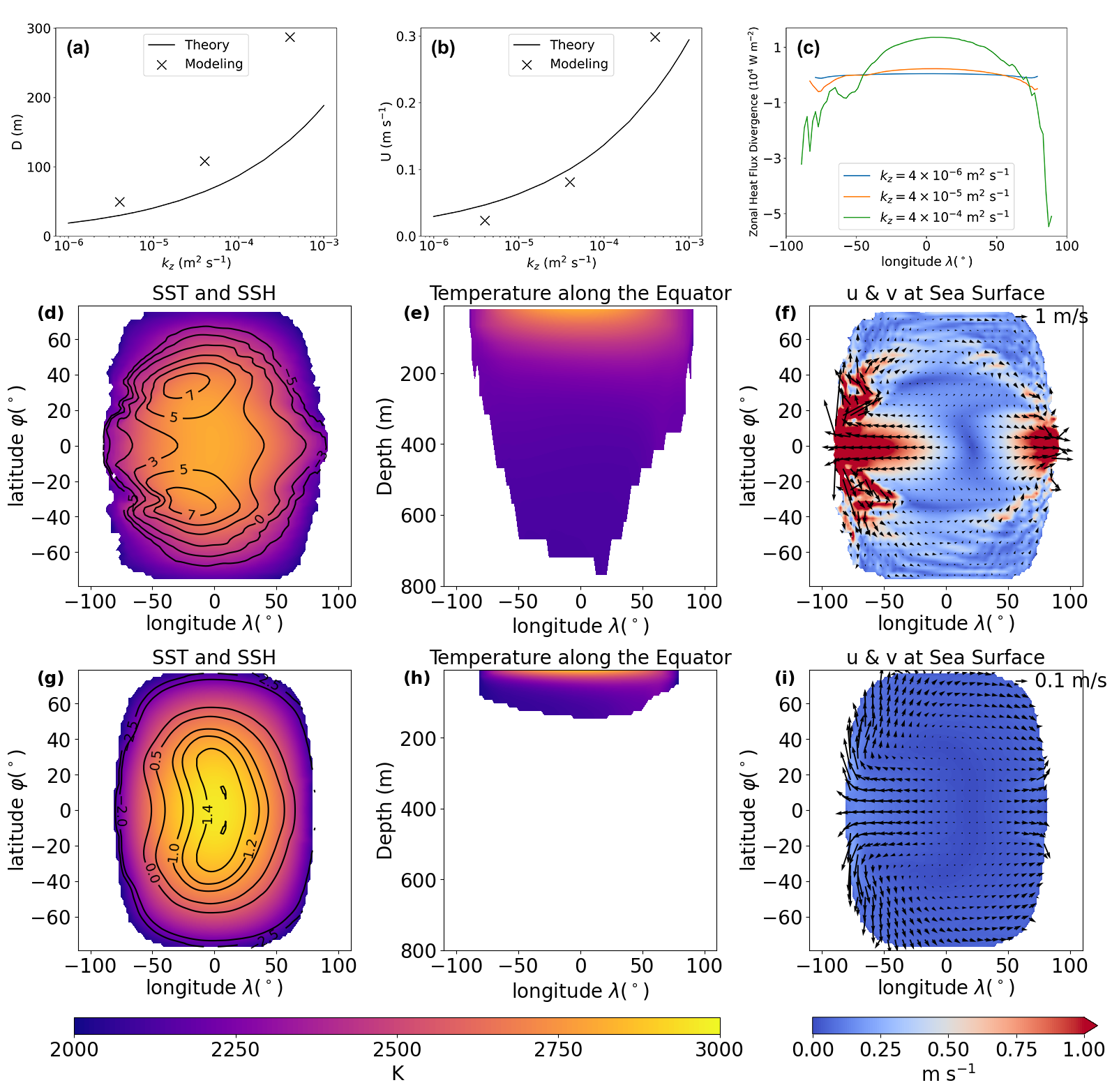}
    \caption{
        Results of the vertical diffusion sensitivity tests. (a) Area-mean ocean depth as a function of the diffusion coefficient, (b) the strength of zonal-mean zonal ocean current, and (c) the divergence of zonal ocean heat transport for the three different diffusion coefficients, $4\times10^{-6}$, $4\times10^{-5}$, and $4\times10^{-4}$ m s$^{-2}$. Note the stellar flux on the planetary surface is in the order of $10^{6}$ W m$^{-2}$, being about two orders larger than the area-mean ocean heat transport. The theoretical lines are from Table 2 in \citet{lai2024b}. \textcolor{black}{(d--f): Sea surface temperature (SST, color-shading in unit of K) and sea surface height (SSH, contour lines in unit of m) (d), lava temperature along the equator (e), and horizontal ocean currents (vectors) and the strength of the currents (color shading) in the experiment with an diffusivity of $4\times10^{-4}$ m s$^{-2}$. (g--i): Same as panels (d--f) but for the experiment of $4\times10^{-6}$ m s$^{-2}$. The contour interval is 2 m in (d) but 0.2 m in (g), and the reference vector is 1.0 m\,s$^{-1}$ in (f) but 0.1 m\,s$^{-1}$ in (i). When the vertical diffusivity is very small, the ocean currents and heat transports are weak and the largest SSH is very close to the substellar point. When the vertical diffusivity increases, the ocean becomes deeper, the ocean currents and associated horizontal and vertical heat transports become stronger, the value of SSH increases, and the largest SSH moves from the equator to the extratropics.}
    }
    \label{fig5}
\end{figure*}

\textcolor{black}{We have also analyzed the ocean circulation in the sensitivity experiments. The three experiments without horizontal currents reveal several common features (Figures A1 \& A2 in the APPENDIX). First, strong currents emerge along the equator, with greater magnitude near the western terminator than the eastern one. Second, a pronounced ``western intensification'' is observed, where currents along the western boundary are significantly stronger than those in the central and eastern regions. Third, large-scale horizontal anticyclonic and cyclonic gyres form in the northern and southern hemispheres, respectively, due to the effect of planetary rotation. Fourth, in the zonal direction—particularly near the equator—strong downwelling occurs near the western and eastern boundaries. Finally, the meridional circulation is characterized by equatorial upwelling (around 30$^{\circ}$S–30$^{\circ}$N) and extratropical downwelling, forming an overturning cell in each hemisphere. While the overall patterns are similar across experiments, the current strengths differ substantially, a variation attributable to the different ocean boundaries determined in the control and sensitivity runs.}

Previous scaling analysis by \citet{lai2024b} showed that in the \textcolor{black}{thermal-driven}, rotation-dominated regime where quasi-geostrophic balance holds, the depth of lava ocean is controlled by a balance between upward advection of heat (or buoyancy) and downward diffusion. This advection-diffusion balance \textcolor{black}{in quasi-geostrophic system} yields a characteristic depth scale given by:
\begin{equation}
D\cong\left(\frac{k_{z}f^{2}L}{\beta\,\Delta\,b}\right)^{\frac{1}{3}}
\label{eq:depth}
\end{equation}
where $k_{z}$ is vertical diffusivity, $f$ is the Coriolis parameter, $\beta$ is its meridional gradient, $L$ is the horizontal scale of large-scale circulation, and $\Delta b$ is horizontal buoyancy contrast. The latter is approximated as $\Delta b \approx g \alpha \Delta T$, where $g$ is gravity, $\alpha$ is the thermal expansion coefficient, and $\Delta T$ is the horizontal temperature contrast between the substellar point and the ocean edge. \textcolor{black}{The key equations to obtain the above scaling are the linear vorticity equation ($\beta v = f\frac{\partial{w}}{\partial{z}}$), geostrophic balance relation ($f \frac{\partial{u}}{\partial{z}} = -\frac{\partial{\Delta b}}{\partial{y}}$), and the advection-diffusive balance ($w \frac{\partial{\theta}}{\partial{z}} = k_z \frac{\partial{^2\theta}}{\partial{z^2}}$), where $v$ is the large-scale horizontal velocity in the meridional direction, $u$ is the large-scale horizontal velocity in the zonal direction but with the same magnitude of $v$, $w$ is the large-scale vertical velocity, and $\theta$ is the potential temperature of the lava. The corresponding scalings of the three equations are $\beta U\thicksim f\frac{W}{D}$, $f\frac{U}{D} \thicksim \frac{\Delta b}{L}$, and $\frac{W}{D} \thicksim \frac{k_z}{D^2}$, where $L$ is the horizontal scale of the large-scale circulation. Combining these three scalings, one can obtain the scaling for the ocean depth (Equation (1)). For more details, please see section 2.3 (Equations 24-28 as well as Equation 3) of \citet{lai2024b}.}


To evaluate this scaling relationship and test the robustness of our boundary iteration method, we performed two sensitivity experiments by varying the vertical diffusivity $k_z$ by an order of magnitude in both directions: increasing it to $4 \times 10^{-4}$ m$^2$\,s$^{-1}$ and decreasing it to $4 \times 10^{-6}$ m$^2$\,s$^{-1}$. As shown in Figure~\ref{fig5}(a \& e), increasing $k_z$ leads to a deeper ocean, with the mean depth rising from approximately 120 m to 280 m. Conversely, decreasing $k_z$ results in a shallower ocean, with a depth of only 50 m (Figure~\ref{fig5}(a \& h)). These results align well with the theoretical scaling (Equation \ref{eq:depth}), supporting the physical consistency of our simulation method. 

Furthermore, the horizontal velocity increases with increasing vertical diffusivity (Figure~\ref{fig5}(b, f \& i)), also consistent with the predictions from the scaling theory \citep[see Table 2 in][]{lai2024b}. 
\textcolor{black}{When the vertical diffusion is small ($4 \times 10^{-6}$ m$^2$\,s$^{-1}$), the ocean circulation becomes week in both vertical and horizontal directions, so the ocean heat transport is tiny and the largest sea surface height is very close to the substellar point. As the vertical diffusion is large ($4 \times 10^{-4}$ m$^2$\,s$^{-1}$), the strong ocean circulation and heat transport deepen the ocean and make the largest sea surface height move to the extratropics.}

The ocean circulation facilitates horizontal heat transport from the substellar region to the ocean boundaries, which in turn helps expand the ocean. However, this \textcolor{black}{area-mean} oceanic heat transport (Figure~\ref{fig5}(c)) is two to three orders of magnitude weaker than the stellar flux (in the order of $10^6$ W\,m$^{-2}$), indicating that horizontal ocean heat transport plays only a minor role in shaping the lateral extent of the ocean. As a result, the lava ocean remains confined on the dayside \textcolor{black}{in all the three experiments}.

Turbulent eddy diffusivity quantified by $k_z$ is a critical but poorly constrained parameter for exoplanetary oceans. In general, vertical mixing arises from tides and wind stress. However, for tidally locked planets in synchronous rotation, the tidal forcing is steady and thus unlikely to drive significant mixing, despite being about 1000 times stronger than Earth's \citep{lingam2018, si2022, shi2025}. While wind forcing is not explicitly modeled in this study, its effects are implicitly included via the assumed nonzero $k_z$. The value adopted in our control experiment, $4 \times 10^{-5}$ m$^2$\,s$^{-1}$, is comparable to the global average for Earth's oceans. Whether this value is appropriate for magma oceans remains uncertain. Future studies incorporating energy budget diagnostics \citep{yang2017, jansen2023} may help constrain $k_z$ more realistically.

Without vertical (diapycnal) mixing, cold dense water sinking from the ocean boundaries would fill the abyss, leaving only a thin warm layer near the surface. Once the deep ocean is filled with dense water, the horizontal pressure gradient between the hot substellar region and the cold ocean edges would vanish, suppressing further convective instability and halting the overturning circulation altogether \citep{vallis2017}. \textcolor{black}{In other words, vertical mixing is necessary for the ocean circulation, but it is extremely challenging to articulate and calculate it accurately.}


\section{Summary and Discussions} \label{sec:summary}

In this study, we simulate \textcolor{black}{thermal-driven} ocean circulation on 1:1 tide-locked lava worlds using the MITgcm combined with a simple boundary iteration method. Our simulations show that \textcolor{black}{the area-mean ocean depth ranges from about 50 to 300 m}, depending on the magnitude of vertical diffusion. The characteristic speed of ocean currents is on the order of 0.1–1 m\,s$^{-1}$, comparable to values found in Earth’s oceans. In the horizontal direction, the lava ocean remains confined to the dayside hemisphere, as the ocean heat transport is relatively weak compared to the incoming stellar radiation. These findings are consistent with theoretical scaling laws presented in \citet{Kite2016} and \citet{lai2024b} as well as 2D (x–z) ideal simulations by \citet{lai2024a}.

Unlike earlier 2D simulations in Cartesian coordinates, our 3D simulations incorporate the Coriolis force on a spherical domain. As a result, several new features emerge, including a clear asymmetry between the western and eastern sides of the substellar point, large-scale hemispheric gyres in the horizontal circulation, western intensification of boundary currents, and a poleward displacement of the deepest ocean region (Figures~\ref{fig3} and \ref{fig4}). \textcolor{black}{Notably, as the vertical diffusion is moderate or large, the greatest ocean depth occurs at the middle latitudes (such as near $40^\circ$S/N) rather than directly beneath the substellar point. When the vertical diffusion is weak, the greatest ocean depth remains near the substellar point.}

\textcolor{black}{In our control experiments, the maximum ocean depth is at about $40^\circ$S/N, because an intermediate rotation period (0.83 Earth days) is employed in the experiment. If the rotation period is very large (such as several hundreds of Earth days), the result will be similar to that with no rotation, so that the location of the maximum ocean depth should be very close to the equator. If the rotation period is very small (such as one hour in theory), the Coriolis force will be strong and the Rossby deformation radius will be small, so that the location of the maximum depth should be also close to the equator. So, we speculate that the location of the maximum ocean depth will be a non-monotonic function of planetary rotation period. In next step of this program, we will thoroughly examine the dependence of the location of the maximum ocean depth on planetary rotation rate as well as other parameters.}

In this work, only the liquid component of the ocean is modeled, and we assume spatially and temporally uniform values for viscosity and diffusivity. Temperature- and pressure-dependent variations in these parameters are not included, as the standard configuration of MITgcm does not support spatially varying viscosity and diffusivity. Nonetheless, given the relatively shallow ocean depths, we expect that the omission of this complexity does not significantly alter our main conclusions.

As a first step toward global 3D modeling of tide-locked lava ocean dynamics, this study focuses exclusively on thermally driven circulation---driven by intense heating over the substellar region and cooling near the ocean boundaries, which together establish horizontal pressure gradients. We have not included the effects of wind stresses, which are likely important given that surface wind speeds on tidally locked planets can reach several kilometers per second \citep[e.g.,][]{castan2011, nguyen2020, kang2021}. Furthermore, we have limited our sensitivity analyses to variations in initial conditions and the vertical diffusivity (Figures~\ref{fig2} and \ref{fig5}). In the next part of this series of study, we will employ a more advanced method, incorporate wind stress forcing, and conduct a broader set of sensitivity experiments to further explore the dynamics of tide-locked lava oceans.\\


\section*{Data Availability}
The experiments were performed with the ocean model MITgcm, which can be downloaded from https://mitgcm.readthedocs.io/en/latest/index.html. The boundary iteration method and the model output data are stored on https://zenodo.org/records/17223885.

\begin{contribution}
J.Y. lead the project, Z.W. installed the model, T.C. performed the simulations, T.C, Z.W. \& J.Y. plotted and improved the figures, J.Y. wrote the manuscript, Y.L. and W.K. improved the manuscript, and all authors discussed the results.
\end{contribution}

\begin{acknowledgments}
We thank the MIT team for developing the model MITgcm. J.Y. is supported by the Natural Science Foundation of China (NSFC) under grant nos. 42441812 and 42161144011. \textcolor{black}{About 200,000 core computing hours have been used for this project. This corresponds to CO$_2$ emission of about 444 kg, if we assume the power per core is 3.7 W and the carbon emission intensity is 0.6 kg/kWh.}
\end{acknowledgments}

\appendix

\section{Results of the Sensitivity Experiments with Different Initial Conditions}

\begin{figure*}[ht!]
    \centering
    \includegraphics[width=\textwidth]{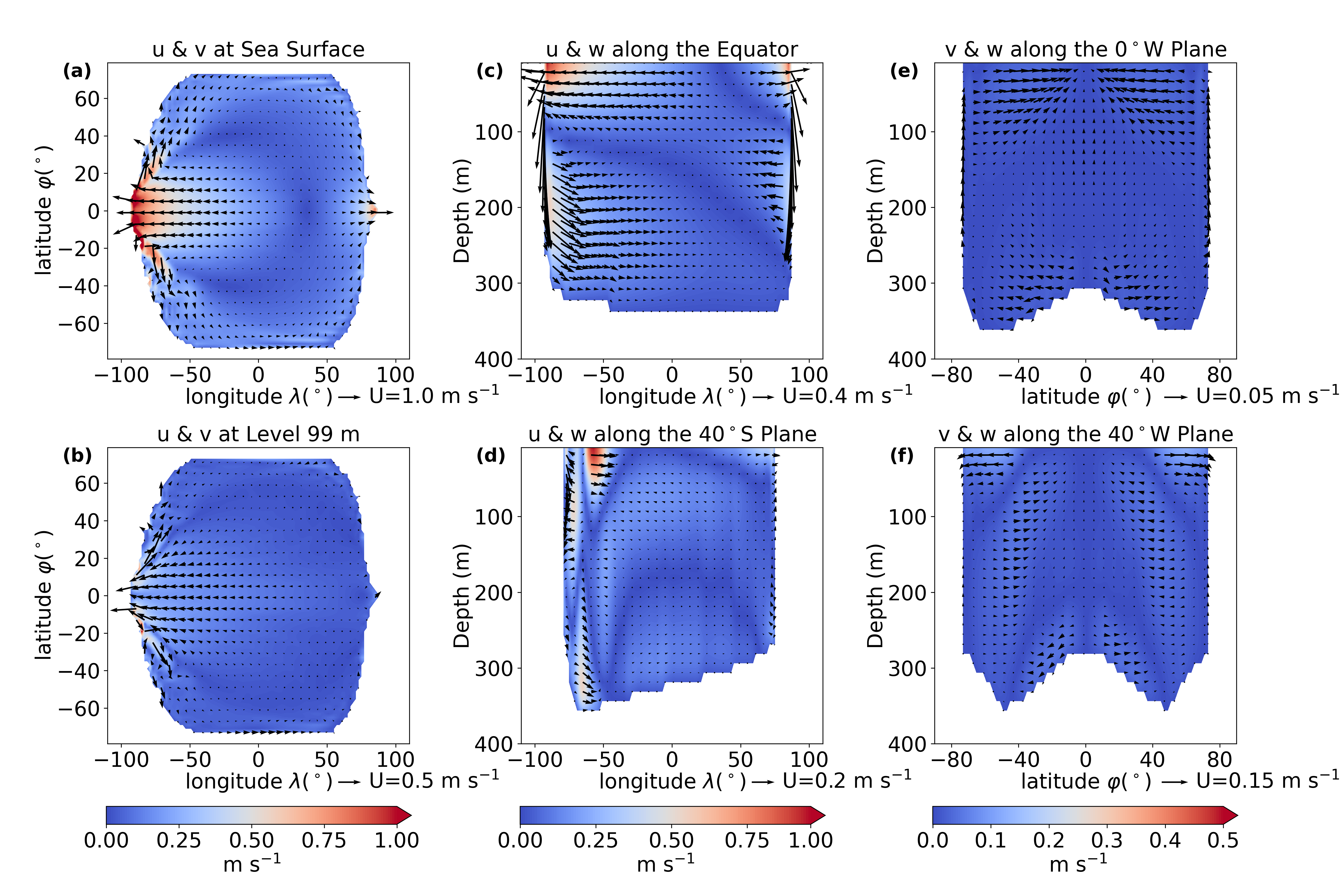}
    \caption{Results of the simulated ocean circulation. Same as Figure 4 in the main text, but for the first sensitivity experiment, within which the initial ocean depth, ocean temperature and ocean current speed are 10 km, 3000 K, and 0, respectively.}
    \label{fig6}
\end{figure*}

\begin{figure*}[ht!]
    \centering
    \includegraphics[width=\textwidth]{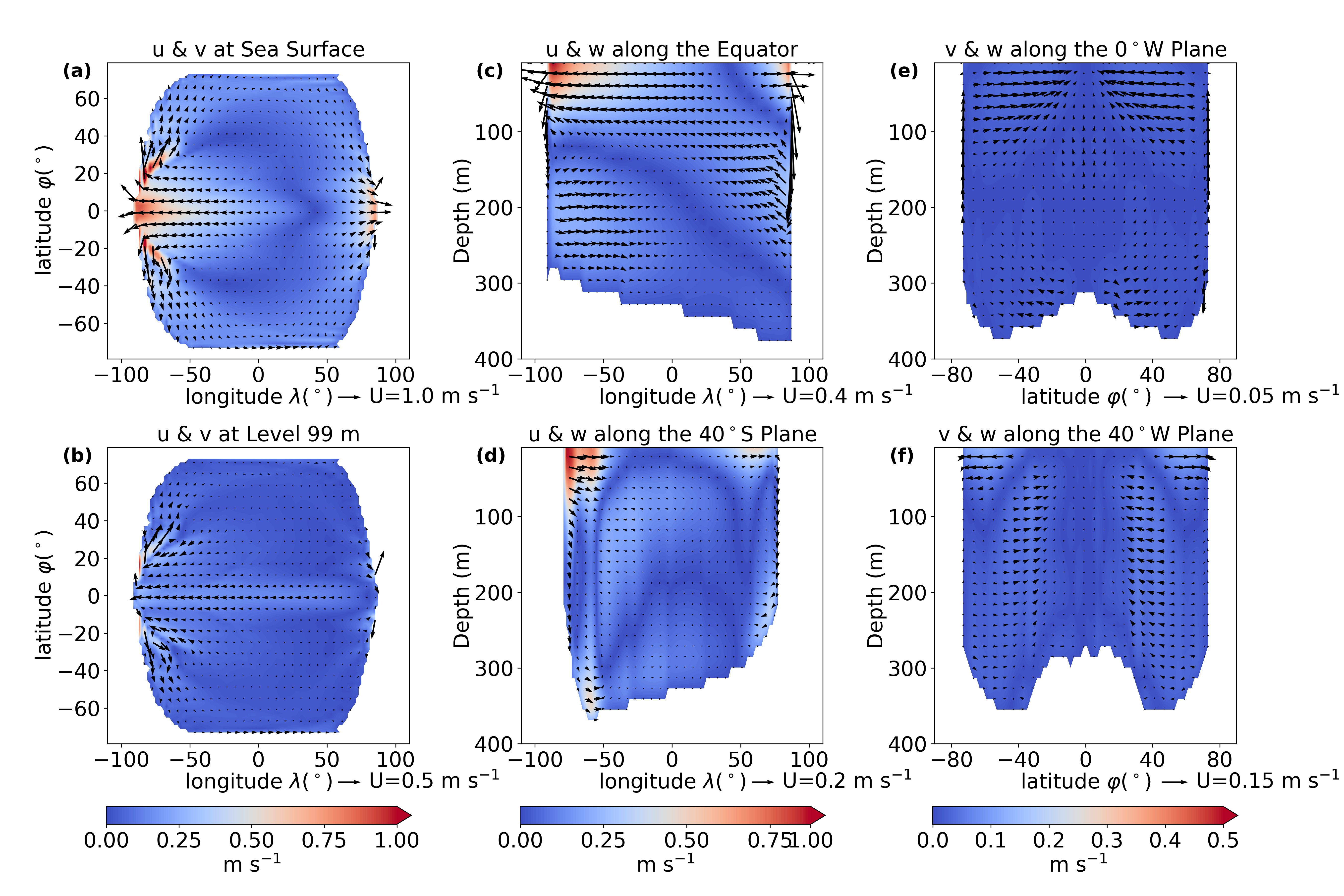}
    \caption{Results of the simulated ocean circulation. Same as Figure 4 in the main text, but for the second sensitivity experiment, within which the initial ocean depth, ocean temperature and ocean current speed are 2 km, 273 K, and 0, respectively.}
    \label{fig7}
\end{figure*}

\bibliography{refs}

@article{showman2008,
  title = {Atmospheric {{Circulation}} of {{Hot Jupiters}}: {{Three-dimensional Circulation Models}} of {{HD}} 209458b and {{HD}} 189733b with {{Simplified Forcing}}},
  shorttitle = {Atmospheric {{Circulation}} of {{Hot Jupiters}}},
  author = {Showman, Adam P. and Cooper, Curtis S. and Fortney, Jonathan J. and Marley, Mark S.},
  year = {2008},
  month = jul,
  journal = {The Astrophysical Journal},
  volume = {682},
  number = {1},
  pages = {559},
  publisher = {IOP Publishing},
  issn = {0004-637X},
  doi = {10.1086/589325},
  urldate = {2025-07-04},
  langid = {english}
}

@article{showman2013,
  title={Atmospheric circulation of terrestrial exoplanets},
  author={Showman, Adam P and Wordsworth, Robin D and Merlis, Timothy M and Kaspi, Yohai},
  journal={Comparative climatology of terrestrial planets},
  volume={1},
  pages={277--326},
  year={2013},
  doi = {10.2458/azu_uapress_9780816530595-ch12},
  publisher={University of Arizona Press, Tuscon, AZ}
}

@article{SchaeferandFegley2009,
  title = {Chemistry of Silicate atmospheres of evaporating super-Earths},
  author = {Schaefer, L. and Fegley JR., E.},
  year = {2009},
  month = oct,
  journal = {The Astrophysical Journal},
  volume = {703},
  pages = {L113--L117},
  doi = {10.1088/0004-637X/703/2/L113},
  urldate = {2009-10-01},
}

@article{nguyen2020,
  title = {Modelling the Atmosphere of Lava Planet {{K2-141b}}: Implications for Low- and High-Resolution Spectroscopy},
  shorttitle = {Modelling the Atmosphere of Lava Planet {{K2-141b}}},
  author = {Nguyen, T Giang and Cowan, Nicolas B and Banerjee, Agnibha and Moores, John E},
  year = {2020},
  month = nov,
  journal = {Monthly Notices of the Royal Astronomical Society},
  volume = {499},
  number = {4},
  pages = {4605--4612},
  issn = {0035-8711},
  doi = {10.1093/mnras/staa2487},
  urldate = {2025-07-04},
}

@article{kang2021,
  title = {Escaping {{Outflows}} from {{Disintegrating Exoplanets}}: {{Day-side}} versus {{Night-side Escape}}},
  shorttitle = {Escaping {{Outflows}} from {{Disintegrating Exoplanets}}},
  author = {Kang, Wanying and Ding, Feng and Wordsworth, Robin and Seager, Sara},
  year = {2021},
  month = jan,
  journal = {The Astrophysical Journal},
  volume = {906},
  number = {2},
  pages = {67},
  publisher = {The American Astronomical Society},
  issn = {0004-637X},
  doi = {10.3847/1538-4357/abcaa7},
  urldate = {2025-05-06},
  langid = {english}
}

@article{zeng2021,
  title = {Oceanic {{Superrotation}} on {{Tidally Locked Planets}}},
  author = {Zeng, Yaoxuan and Yang, Jun},
  year = {2021},
  month = mar,
  journal = {The Astrophysical Journal},
  volume = {909},
  number = {2},
  pages = {172},
  publisher = {The American Astronomical Society},
  issn = {0004-637X},
  doi = {10.3847/1538-4357/abe12f},
  urldate = {2025-06-11},
  langid = {english}
}

@book{stewart2008,
  author = {Stewart, Robert H.},
  title = {Introduction to Physical Oceanography},
  year = {2008},
  publisher = {Orange Grove},
  address = {Tallahassee, FL},
  isbn = {1616100451},
  url = {https://oceanworld.tamu.edu/resources/ocng_textbook/contents.html},
  note = {Free online edition available from Texas A\&M University}
}

@article{jansen2023,
  title = {Energetic Constraints on Ocean Circulations of Icy Ocean Worlds},
  author = {Jansen, Malte F. and Kang, Wanying and Kite, Edwin S. and Zeng, Yaoxuan},
  year = {2023},
  month = jun,
  journal = {The Planetary Science Journal},
  volume = {4},
  number = {6},
  pages = {117},
  publisher = {The American Astronomical Society},
  doi = {10.3847/PSJ/acda95},
}

@article{Shi2025,
  title = {Ocean Tides on Asynchronously Rotating Planets Orbiting Low-mass Stars},
  author = {Shi, Jiaru and Abbot, Dorian S. and Liu, Yonggang and Kang, Wanying and Lin, Yufeng},
  year = {2025},
  journal = {The Astrophysical Journal},
  volume = {989},
  number = {139},
  pages = {1--13},
  publisher = {the American Astronomical Society},
  doi = {10.3847/1538-4357/adeb83},
  urldate = {2025-08-11},
}

@article{Katsuraetal2010,
  title = {Adiabatic temperature profile in the mantle},
  author = {Katsura, T. and Yoneda, A. and Yamazaki, D. and Yoshino, T. and Ito, E.},
  year = {2010},
  journal = {Physics of the Earth and Planetary Interiors},
  volume = {183},
  number = {1--2},
  pages = {212--218},
  doi = {https://doi.org/10.1016/j.pepi.2010.07.001},
  urldate = {2025-08-11},
}

@article{Sakamakietal2010,
  title = {Density of dry peridotite magma at high pressure using an X-ray absorption method},
  author = {Sakamaki, T. and Ohtani, E. and Urakawa, S. and Suzuki, A. and Katayama, Y.},
  year = {2010},
  journal = {American Mineralogist},
  volume = {95},
  pages = {144--147},
  publisher = {the American Astronomical Society},
  doi = {https://doi.org/10.2138/am.2010.3143},
}

@article{yang2017,
  title = {Persistence of a Freshwater Surface Ocean after a Snowball {{Earth}}},
  author = {Yang, Jun and Jansen, Malte F. and Macdonald, Francis A. and Abbot, Dorian S.},
  year = {2017},
  month = jul,
  journal = {Geology},
  volume = {45},
  number = {7},
  pages = {615--618},
  publisher = {GeoScienceWorld},
  issn = {0091-7613},
  doi = {10.1130/G38920.1},
  urldate = {2025-06-11},
  langid = {english}
}

@article{lingam2018,
  title = {Implications of {{Tides}} for {{Life}} on {{Exoplanets}}},
  author = {Lingam, Manasvi and Loeb, Abraham},
  year = {2018},
  month = jul,
  journal = {Astrobiology},
  volume = {18},
  number = {7},
  pages = {967--982},
  publisher = {Mary Ann Liebert, Inc., publishers},
  issn = {1531-1074},
  doi = {10.1089/ast.2017.1718},
  urldate = {2025-06-11},
  langid = {american}
}

@article{si2022,
  title = {Planetary Climate under Extremely High Vertical Diffusivity},
  author = {Si, Yidongfang and Yang, Jun and Liu, Yonggang},
  year = {2022},
  month = feb,
  journal = {Astronomy \& Astrophysics},
  volume = {658},
  pages = {A33},
  publisher = {EDP Sciences},
  issn = {0004-6361, 1432-0746},
  doi = {10.1051/0004-6361/202140778},
  urldate = {2025-06-11},
  copyright = {{\copyright} ESO 2022},
  langid = {english}
}

@article{stommel1948,
  author = {Stommel, Henry},
  title = {The Westward Intensification of Wind-Driven Ocean Currents},
  journal = {Transactions of the American Geophysical Union},
  year = {1948},
  volume = {29},
  pages = {202--206},
  doi = {10.1029/TR029i002p00202},
}

@article{munk1950,
  author = {Munk, Walter},
  title = {On the Wind-Driven Ocean Circulation},
  journal = {Journal of Meteorology},
  year = {1950},
  volume = {7},
  pages = {80--93},
  doi = {10.1175/1520-0469(1950)007<0080:OTWDOC>2.0.CO;2},
}

@article{castan2011,
  title = {{{ATMOSPHERES OF HOT SUPER-EARTHS}}},
  author = {Castan, Thibaut and Menou, Kristen},
  year = {2011},
  month = nov,
  journal = {The Astrophysical Journal Letters},
  volume = {743},
  number = {2},
  pages = {L36},
  publisher = {The American Astronomical Society},
  issn = {2041-8205},
  doi = {10.1088/2041-8205/743/2/L36},
  urldate = {2025-06-04},
  langid = {english}
}

@article{boyd2014,
  title = {Geometrical Effects on Western Intensification of Wind-Driven Ocean Currents: {{The}} Rotated-Channel {{Stommel}} Model, Coastal Orientation, and Curvature},
  shorttitle = {Geometrical Effects on Western Intensification of Wind-Driven Ocean Currents},
  author = {Boyd, John P. and Sanjaya, Edwin},
  year = {2014},
  month = mar,
  journal = {Dynamics of Atmospheres and Oceans},
  volume = {65},
  pages = {17--38},
  issn = {0377-0265},
  doi = {10.1016/j.dynatmoce.2013.10.003},
  urldate = {2025-06-04}
}

@article{zhang2022,
  title = {Internal Dynamics of Magma Ocean and Its Linkage to Atmospheres},
  author = {Zhang, Yizhuo and Zhang, Nan and Tian, Meng},
  year = {2022},
  month = aug,
  journal = {Acta Geochimica},
  volume = {41},
  number = {4},
  pages = {568--591},
  issn = {2365-7499},
  doi = {10.1007/s11631-021-00514-x},
  langid = {american}
}

@article{sun2020,
  title = {Physical State of an Early Magma Ocean Constrained by the Thermodynamics and Viscosity of Iron Silicate Liquid},
  author = {Sun, Yicheng and Zhou, Huiqun and Liu, Xiandong and Yin, Kun and Lu, Xiancai},
  year = {2020},
  month = dec,
  journal = {Earth and Planetary Science Letters},
  volume = {551},
  pages = {116556},
  issn = {0012-821X},
  doi = {10.1016/j.epsl.2020.116556},
  urldate = {2025-06-04}
}

@article{dingwell2004,
  title = {Viscosity of Peridotite Liquid},
  author = {Dingwell, D. B. and Courtial, P. and Giordano, D. and Nichols, A. R. L.},
  year = {2004},
  month = sep,
  journal = {Earth and Planetary Science Letters},
  volume = {226},
  number = {1},
  pages = {127--138},
  issn = {0012-821X},
  doi = {10.1016/j.epsl.2004.07.017},
  urldate = {2025-06-04}
}

@article{dumusque2014,
  title = {{{THE KEPLER-10 PLANETARY SYSTEM REVISITED BY HARPS-N}}: {{A HOT ROCKY WORLD AND A SOLID NEPTUNE-MASS PLANET}}*},
  shorttitle = {{{THE KEPLER-10 PLANETARY SYSTEM REVISITED BY HARPS-N}}},
  author = {Dumusque, Xavier and Bonomo, Aldo S. and Haywood, Rapha{\"e}lle D. and Malavolta, Luca and S{\'e}gransan, Damien and Buchhave, Lars A. and Cameron, Andrew Collier and Latham, David W. and Molinari, Emilio and Pepe, Francesco and Udry, St{\'e}phane and Charbonneau, David and Cosentino, Rosario and Dressing, Courtney D. and Figueira, Pedro and Fiorenzano, Aldo F. M. and Gettel, Sara and Harutyunyan, Avet and Horne, Keith and {Lopez-Morales}, Mercedes and Lovis, Christophe and Mayor, Michel and Micela, Giusi and Motalebi, Fatemeh and Nascimbeni, Valerio and Phillips, David F. and Piotto, Giampaolo and Pollacco, Don and Queloz, Didier and Rice, Ken and Sasselov, Dimitar and Sozzetti, Alessandro and Szentgyorgyi, Andrew and Watson, Chris},
  year = {2014},
  month = jun,
  journal = {The Astrophysical Journal},
  volume = {789},
  number = {2},
  pages = {154},
  publisher = {The American Astronomical Society},
  issn = {0004-637X},
  doi = {10.1088/0004-637X/789/2/154},
  urldate = {2025-06-04},
  langid = {english}
}

@article{batalha2011,
  title = {{{KEPLER}}'{{S FIRST ROCKY PLANET}}: {{KEPLER-10b}}*},
  shorttitle = {{{KEPLER}}'{{S FIRST ROCKY PLANET}}},
  author = {Batalha, Natalie M. and Borucki, William J. and Bryson, Stephen T. and Buchhave, Lars A. and Caldwell, Douglas A. and {Christensen-Dalsgaard}, J{\o}rgen and Ciardi, David and Dunham, Edward W. and Fressin, Francois and Gautier, Thomas N. and Gilliland, Ronald L. and Haas, Michael R. and Howell, Steve B. and Jenkins, Jon M. and Kjeldsen, Hans and Koch, David G. and Latham, David W. and Lissauer, Jack J. and Marcy, Geoffrey W. and Rowe, Jason F. and Sasselov, Dimitar D. and Seager, Sara and Steffen, Jason H. and Torres, Guillermo and Basri, Gibor S. and Brown, Timothy M. and Charbonneau, David and Christiansen, Jessie and Clarke, Bruce and Cochran, William D. and Dupree, Andrea and Fabrycky, Daniel C. and Fischer, Debra and Ford, Eric B. and Fortney, Jonathan and Girouard, Forrest R. and Holman, Matthew J. and Johnson, John and Isaacson, Howard and Klaus, Todd C. and Machalek, Pavel and Moorehead, Althea V. and Morehead, Robert C. and Ragozzine, Darin and Tenenbaum, Peter and Twicken, Joseph and Quinn, Samuel and VanCleve, Jeffrey and Walkowicz, Lucianne M. and Welsh, William F. and Devore, Edna and Gould, Alan},
  year = {2011},
  month = feb,
  journal = {The Astrophysical Journal},
  volume = {729},
  number = {1},
  pages = {27},
  publisher = {The American Astronomical Society},
  issn = {0004-637X},
  doi = {10.1088/0004-637X/729/1/27},
  urldate = {2025-06-04},
  langid = {english}
}

@article{marshall1997_b,
  title = {Hydrostatic, Quasi-Hydrostatic, and Nonhydrostatic Ocean Modeling},
  author = {Marshall, John and Hill, Chris and Perelman, Lev and Adcroft, Alistair},
  year = {1997},
  journal = {Journal of Geophysical Research: Oceans},
  volume = {102},
  number = {C3},
  pages = {5733--5752},
  issn = {2156-2202},
  doi = {10.1029/96JC02776},
  urldate = {2025-06-04},
  copyright = {Copyright 1997 by the American Geophysical Union.},
  langid = {english}
}

@article{marshall1997_a,
  title = {A Finite-Volume, Incompressible {{Navier Stokes}} Model for Studies of the Ocean on Parallel Computers},
  author = {Marshall, John and Adcroft, Alistair and Hill, Chris and Perelman, Lev and Heisey, Curt},
  year = {1997},
  journal = {Journal of Geophysical Research: Oceans},
  volume = {102},
  number = {C3},
  pages = {5753--5766},
  issn = {2156-2202},
  doi = {10.1029/96JC02775},
  urldate = {2025-06-04},
  copyright = {Copyright 1997 by the American Geophysical Union.},
  langid = {english}
}

@book{vallis2017,
  author    = {Geoffrey K. Vallis},
  title     = {Atmospheric and Oceanic Fluid Dynamics: Fundamentals and Large-Scale Circulation},
  edition   = {2nd},
  publisher = {Cambridge University Press},
  year      = {2017},
  isbn      = {9781107065508},
  url       = {
https://doi.org/10.1017/CBO9780511790447}
}

@article{Boukare2022,
  title = {Deep {{Two-phase}}, {{Hemispherical Magma Oceans}} on {{Lava Planets}}},
  author = {Boukar{\'e}, Charles-{\'E}douard and Cowan, Nicolas B. and Badro, James},
  year = {2022},
  month = sep,
  journal = {The Astrophysical Journal},
  volume = {936},
  number = {2},
  pages = {148},
  publisher = {The American Astronomical Society},
  issn = {0004-637X},
  doi = {10.3847/1538-4357/ac8792},
  urldate = {2025-06-04},
  langid = {english}
}

@article{leger2009,
  title = {Transiting Exoplanets from the {{CoRoT}} Space Mission. {{VIII}}. {{CoRoT-7b}}: The First Super-{{Earth}} with Measured Radius},
  shorttitle = {Transiting Exoplanets from the {{CoRoT}} Space Mission. {{VIII}}. {{CoRoT-7b}}},
  author = {L{\'e}ger, A. and Rouan, D. and Schneider, J. and Barge, P. and Fridlund, M. and Samuel, B. and Ollivier, M. and Guenther, E. and Deleuil, M. and Deeg, H. J. and Auvergne, M. and Alonso, R. and Aigrain, S. and Alapini, A. and Almenara, J. M. and Baglin, A. and Barbieri, M. and Bruntt, H. and Bord{\'e}, P. and Bouchy, F. and Cabrera, J. and Catala, C. and Carone, L. and Carpano, S. and Csizmadia, {\relax Sz}. and Dvorak, R. and Erikson, A. and {Ferraz-Mello}, S. and Foing, B. and Fressin, F. and Gandolfi, D. and Gillon, M. and Gondoin, {\relax Ph}. and Grasset, O. and Guillot, T. and Hatzes, A. and H{\'e}brard, G. and Jorda, L. and Lammer, H. and Llebaria, A. and Loeillet, B. and Mayor, M. and Mazeh, T. and Moutou, C. and P{\"a}tzold, M. and Pont, F. and Queloz, D. and Rauer, H. and Renner, S. and Samadi, R. and Shporer, A. and Sotin, {\relax Ch}. and Tingley, B. and Wuchterl, G. and Adda, M. and Agogu, P. and Appourchaux, T. and Ballans, H. and Baron, P. and Beaufort, T. and Bellenger, R. and Berlin, R. and Bernardi, P. and Blouin, D. and Baudin, F. and Bodin, P. and Boisnard, L. and Boit, L. and Bonneau, F. and Borzeix, S. and Briet, R. and Buey, J. -T. and Butler, B. and Cailleau, D. and Cautain, R. and Chabaud, P. -Y. and Chaintreuil, S. and Chiavassa, F. and Costes, V. and Cuna Parrho, V. and {de Oliveira Fialho}, F. and Decaudin, M. and Defise, J. -M. and Djalal, S. and Epstein, G. and Exil, G. -E. and Faur{\'e}, C. and Fenouillet, T. and Gaboriaud, A. and Gallic, A. and Gamet, P. and Gavalda, P. and Grolleau, E. and Gruneisen, R. and Gueguen, L. and Guis, V. and Guivarc'h, V. and Guterman, P. and Hallouard, D. and Hasiba, J. and Heuripeau, F. and Huntzinger, G. and Hustaix, H. and Imad, C. and Imbert, C. and Johlander, B. and Jouret, M. and Journoud, P. and Karioty, F. and Kerjean, L. and Lafaille, V. and Lafond, L. and {Lam-Trong}, T. and Landiech, P. and Lapeyrere, V. and Larqu{\'e}, T. and Laudet, P. and Lautier, N. and Lecann, H. and Lefevre, L. and Leruyet, B. and Levacher, P. and Magnan, A. and Mazy, E. and Mertens, F. and Mesnager, J. -M. and Meunier, J. -C. and Michel, J. -P. and Monjoin, W. and Naudet, D. and {Nguyen-Kim}, K. and Orcesi, J. -L. and Ottacher, H. and Perez, R. and Peter, G. and Plasson, P. and Plesseria, J. -Y. and Pontet, B. and Pradines, A. and Quentin, C. and Reynaud, J. -L. and Rolland, G. and Rollenhagen, F. and Romagnan, R. and Russ, N. and Schmidt, R. and Schwartz, N. and Sebbag, I. and Sedes, G. and Smit, H. and Steller, M. B. and Sunter, W. and Surace, C. and Tello, M. and Tiph{\`e}ne, D. and Toulouse, P. and Ulmer, B. and Vandermarcq, O. and Vergnault, E. and Vuillemin, A. and Zanatta, P.},
  year = {2009},
  month = oct,
  journal = {Astronomy and Astrophysics},
  volume = {506},
  pages = {287--302},
  issn = {0004-6361},
  doi = {10.1051/0004-6361/200911933},
  urldate = {2025-06-11}
}

@article{leger2011,
title = {The extreme physical properties of the CoRoT-7b super-Earth},
journal = {Icarus},
volume = {213},
number = {1},
pages = {1-11},
year = {2011},
issn = {0019-1035},
doi = {https://doi.org/10.1016/j.icarus.2011.02.004},
url = {https://www.sciencedirect.com/science/article/pii/S0019103511000534},
author = {A. Léger and O. Grasset and B. Fegley and F. Codron and A.F. Albarede and P. Barge and R. Barnes and P. Cance and S. Carpy and F. Catalano and C. Cavarroc and O. Demangeon and S. Ferraz-Mello and P. Gabor and J.-M. Grießmeier and J. Leibacher and G. Libourel and A.-S. Maurin and S.N. Raymond and D. Rouan and B. Samuel and L. Schaefer and J. Schneider and P.A. Schuller and F. Selsis and C. Sotin}
}

@article{Meier2023,
title = {Interior dynamics of super-Earth 55 Cancri e},
journal = {A\&A},
volume = {678},
number = {A29},
pages = {18},
year = {2023},
doi = {https://doi.org/10.1051/0004-6361/202346950},
author = {Tobias G. Meier and Dan J. Bower and Tim Lichtenberg and Mark Hammond and Paul J. Tackley}
}

@article{chao2021,
title = {Lava worlds: From early earth to exoplanets},
journal = {Geochemistry},
volume = {81},
number = {2},
pages = {125735},
year = {2021},
issn = {0009-2819},
doi = {https://doi.org/10.1016/j.chemer.2020.125735},
url = {https://www.sciencedirect.com/science/article/pii/S000928192030146X},
author = {Keng-Hsien Chao and Rebecca deGraffenried and Mackenzie Lach and William Nelson and Kelly Truax and Eric Gaidos}
}

@article{Kite2016,
  title = {{{ATMOSPHERE-INTERIOR EXCHANGE ON HOT}}, {{ROCKY EXOPLANETS}}},
  author = {Kite, Edwin S. and Jr, Bruce Fegley and Schaefer, Laura and Gaidos, Eric},
  year = {2016},
  month = sep,
  journal = {The Astrophysical Journal},
  volume = {828},
  number = {2},
  pages = {80},
  publisher = {The American Astronomical Society},
  issn = {0004-637X},
  doi = {10.3847/0004-637X/828/2/80},
  urldate = {2025-05-06},
  langid = {english}
}

@article{lai2024a,
  title = {Ocean {{Circulation}} on {{Tide-locked Lava Worlds}}. {{I}}. {{An Idealized 2D Numerical Model}}},
  author = {Lai, Yanhong and Yang, Jun and Kang, Wanying},
  year = {2024a},
  month = sep,
  journal = {The Planetary Science Journal},
  volume = {5},
  number = {9},
  pages = {204},
  publisher = {IOP Publishing},
  issn = {2632-3338},
  doi = {10.3847/PSJ/ad7111},
  urldate = {2025-06-04},
  langid = {english}
}

@article{lai2024b,
  title = {Ocean {{Circulation}} on {{Tide-locked Lava Worlds}}. {{II}}. {{Scalings}}},
  author = {Lai, Yanhong and Kang, Wanying and Yang, Jun},
  year = {2024b},
  month = sep,
  journal = {The Planetary Science Journal},
  volume = {5},
  number = {9},
  pages = {205},
  publisher = {The American Astronomical Society},
  doi = {10.3847/PSJ/ad70b4},
  langid = {american}
}
\bibliographystyle{aasjournalv7}

\end{document}